\documentclass[twocolumn,pra,10pt,showpacs]{revtex4}
\usepackage{graphicx,amsmath,amssymb}

\newcommand{\beq}{\begin{equation}}
\newcommand{\eeq}{\end{equation}}
\newcommand{\beqa}{\begin{eqnarray}}
\newcommand{\eeqa}{\end{eqnarray}}
\newcommand{\ket}[1]{| #1 \rangle}
\newcommand{\bra}[1]{\langle #1 |}

\begin{document}


\title{Concurrence classes for an arbitrary multi-qubit  state based on positive operator valued measure}

\author{Hoshang Heydari}
\email{hoshang@imit.kth.se} \affiliation{Institute of Quantum
Science, Nihon University, 1-8 Kanda-Surugadai, Chiyoda-ku, Tokyo
101- 8308, Japan}

\date{\today}

\begin{abstract}
In this paper, we propose concurrence classes  for an arbitrary
multi-qubit state based on orthogonal complement of a positive
operator valued measure, or POVM in short, on quantum phase. In
particular, we construct concurrence for an arbitrary two-qubit
state and concurrence classes for the three- and four-qubit
states. And finally, we construct  $W^{m}$ and $GHZ^{m}$ class
concurrences for multi-qubit states. The unique structure of our
POVM enables us to distinguish different concurrence classes for
multi-qubit states.
\end{abstract}


\maketitle

\section{Introduction}
Entanglement is an interesting feature of quantum theory which in
recent years attract many researcher to quantify, classify, and to
investigate its useful properties. Entanglement has already some
applications such as quantum teleportation and quantum key
distribution, and it surely will arrive new applications for this
fascinating quantum phenomenon. For instance, multipartite
entanglement has a capacity to offer new unimaginable applications
in emerging fields of quantum information and quantum computation.
One of widely used measures of entanglement for a pair of qubits
is  the concurrence that gives an analytic formula for the
entanglement of formation \cite{Bennett96,Wootters98}. In recent
years, there have been made some proposals to generalize this
measure into a general bipartite state, e.g., Uhlmann
\cite{Uhlmann00} has generalized the concept of concurrence by
considering arbitrary conjugation, than Audenaert \emph{et
al.}\cite{Audenaert} generalized this formula in spirit of
Uhlmann's work, by defining a concurrence vector for pure state.
 Moreover, Gerjuoy \cite{Gerjuoy}
and Albeverio and Fei \cite{Albeverio} gave an explicit expression
in terms of coefficient of a general pure bipartite state.
Therefore, it could be interesting to try to generalized this
measure from bipartite to multipartite system, see Ref.
\cite{Bhaktavatsala,Akhtarshenas,Oster,Wang}. An application of
concurrence for a physically realizable state such as BCS state
can be found in Ref. \cite{Martin}. Quantifying entanglement of
multipartite states has been discussed in
\cite{Lewen00,Vedral97,Werner89,Hor00,Acin01,Bennett96a,Dur99,
ECKERT02,Dur00,Eisert01,Verst03,Pan}. In \cite{Hosh3,Hosh4} we
have proposed a degree of entanglement for a general pure
multipartite state, based on the POVM on quantum phase. In this
paper, we will define concurrence for an arbitrary two-qubit state
based on orthogonal complement of our POVM. From our POVM we will
construct an operator that can be seen as a tiled operation acting
on the density operator. Moreover, we will define concurrences for
different classes of arbitrary three- and four-qubit  states. And
finally, we will generalize our result into an arbitrary
multi-qubit  state. The structure of our POVM enables us to detect
and to define different concurrence classes for multi-qubit
states. The definition of concurrence is based on an analogy with
bipartite state.
For multi-qubit states, the $W^{m}$ class concurrences are
invariant under stochastic local quantum operation and classical
communication(SLOCC) \cite{Dur00}. Furthermore, all homogeneous
positive functions of pure states that are invariant under
determinant-one SLOCC operations are entanglement monotones
\cite{Verst03}. However, invariance under SLOCC  for the $W^{m}$
class concurrence for general multipartite states need deeper
investigation. It is worth mentioning  that Uhlmann
\cite{Uhlmann00} has shown that entanglement monotones for
concurrence are related to antilinear operators. However, the
$GHZ^{m}$ class concurrences for multipartite states need
optimization over all local unitary operations. Classification of
multipartite states has been discussed in
\cite{Verst,Oster,Miyake,Miyake04,Mintert,Wang}. For example, F.
Verstraete \emph{et al}. \cite{Verst} have considered a single
copy of a pure four-partite state of qubits and investigated its
behavior under SLOCC, which gave a classification of all different
classes of pure states of four qubits. They have also shown that
there exist nine families of states corresponding to nine
different ways of entangling four qubits. A. Osterloh and J.
Siewert \cite{Oster} have
 constructed entanglement measures for pure
states of multipartite qubit systems. The key element of their
approach is an antilinear operator that they called comb. For
qubits, the combs are invariant under the action of the special
linear group. They have also discussed inequivalent types of
genuine four-qubit entanglement, and found three types of
entanglement for these states. This result coincides with our
classification, where in section \ref{Sec4} we construct three
types of concurrence classes for four-qubit states.  A. Miyake
\cite{Miyake}, has also discussed classification of multipartite
states in entanglement classes based on the hyper-determinant. He
shown that two states belong to the same class if they are
interconvertible under SLOCC. Moreover, the only paper that
addressed the classification of higher-dimensional multipartite
states is the paper by A. Miyake and F. Verstraete
\cite{Miyake04}, where they have classified multipartite entangled
states in the $ 2\times2 \times n$ quantum systems for (n $\geq$
4). They have shown that there exist nine essentially different
classes of states, and they give rise to a five-graded partially
ordered structure, including GHZ class and W class of 3 qubits. F.
Mintert \emph{et al}. \cite{Mintert} have proposed generalizations
of concurrence for multi-partite quantum systems that can
distinguish distinct quantum correlations. However, their
construction is not similar to our concurrence classes, since we
can distinguish these classes based on joint phases of the
orthogonal complement of our POVM by construction. Finally, A. M.
Wang \cite{Wang} has proposed two classes of the generalized
concurrence vectors of the multipartite systems consisting of
qubits. Our classification is similar to Wang's classification of
multipartite state. However, the advantage of our method is that
our POVM can distinguish these concurrence classes without prior
information about inequivalence  of these classes under local
quantum operation and classical communication (LOCC).
Let us denote a general, multipartite quantum system with $m$
subsystems by
$\mathcal{Q}=\mathcal{Q}_{m}(N_{1},N_{2},\ldots,N_{m})$ $
=\mathcal{Q}_{1}\mathcal{Q}_{2}\cdots\mathcal{Q}_{m}$, consisting
of a state $
\ket{\Psi}=\sum^{N_{1}}_{k_{1}=1}\cdots\sum^{N_{m}}_{k_{m}=1}
\alpha_{k_{1},\ldots,k_{m}} \ket{k_{1},\ldots,k_{m}} $ and  $
\ket{\Psi^{*}}=\sum^{N_{1}}_{k_{1}=1}\cdots\sum^{N_{m}}_{k_{m}=1}
\alpha^{*}_{k_{1},\ldots,k_{m}} \ket{k_{1},\ldots,k_{m}} $, the
complex conjugate of $\ket{\Psi}$, let
$\rho_{\mathcal{Q}}=\sum^{\mathrm{N}}_{n=1}p_{n}\ket{\Psi_{n}}\bra{\Psi_{n}}$,
for all $0\leq p_{n}\leq 1$ and $\sum^{\mathrm{N}}_{n=1}p_{n}=1$,
denote a density operator acting on the Hilbert space $
\mathcal{H}_{\mathcal{Q}}=\mathcal{H}_{\mathcal{Q}_{1}}\otimes
\mathcal{H}_{\mathcal{Q}_{2}}\otimes\cdots\otimes\mathcal{H}_{\mathcal{Q}_{m}},
$ where the dimension of the $j$th Hilbert space is given  by
$N_{j}=\dim(\mathcal{H}_{\mathcal{Q}_{j}})$. We are going to use
this notation throughout this paper, i.e., we denote a mixed pair
of qubits by $\mathcal{Q}_{2}(2,2)$. The density operator
$\rho_{\mathcal{Q}}$ is said to be fully separable, which we will
denote by $\rho^{sep}_{\mathcal{Q}}$, with respect to the Hilbert
space decomposition, if it can  be written as $
\rho^{sep}_{\mathcal{Q}}=\sum^\mathrm{N}_{n=1}p_{n}
\bigotimes^m_{j=1}\rho^{n}_{\mathcal{Q}_{j}}$,
$\sum^\mathrm{N}_{n=1}p_{n}=1 $, for some positive integer
$\mathrm{N}$, where $p_{n}$ are positive real numbers and
$\rho^{n}_{\mathcal{Q}_{j}}$ denotes a density operator on Hilbert
space $\mathcal{H}_{\mathcal{Q}_{j}}$. If $\rho^{p}_{\mathcal{Q}}$
represents a pure state, then the quantum system is fully
separable if $\rho^{p}_{\mathcal{Q}}$ can be written as
$\rho^{sep}_{\mathcal{Q}}=\bigotimes^m_{j=1}\rho_{\mathcal{Q}_{j}}$,
where $\rho_{\mathcal{Q}_{j}}$ is a density operator on
$\mathcal{H}_{\mathcal{Q}_{j}}$. If a state is not separable, then
it is called an entangled state. Some of the generic entangled
states are called Bell states and $\mathrm{EPR}$ states.

\section{General definition of POVM on quantum phase }
In this section we will define a general POVM on quantum phase.
This POVM is a set of linear operators
$\Delta(\varphi_{1,2},\ldots,\varphi_{1,N},\varphi_{2,3},\ldots,\varphi_{N-1,N})$
furnishing the probabilities that the measurement of a state
$\rho$ on the Hilbert space $\mathcal{H}$ is given by
\begin{eqnarray}
&&\mathrm{p}(\varphi_{1,2},\ldots,\varphi_{1,N},\varphi_{2,3},\ldots,\varphi_{N-1,N})\\\nonumber
&=&
\mathrm{Tr}(\rho\Delta(\varphi_{1,2},\ldots,\varphi_{1,N},\varphi_{2,3},\ldots,\varphi_{N-1,N})),
\end{eqnarray}
where
$(\varphi_{1,2},\ldots,\varphi_{1,N},\varphi_{2,3},\ldots,\varphi_{N-1,N})$
are the outcomes of the measurement of the quantum  phase, which is
discrete and binary. This POVM satisfies the following properties,
$\Delta(\varphi_{1,2},\ldots,\varphi_{1,N},\varphi_{2,3},\ldots,\varphi_{N-1,N})$
is self-adjoint, is positive, and  is normalized, i.e.,
\begin{eqnarray}&&\overbrace{\int_{2\pi}\cdots
\int_{2\pi}}^{N(N-1)/2}d\varphi_{1,2}\cdots
d\varphi_{1,N}d\varphi_{2,3}\\\nonumber&&\cdots d\varphi_{N-1,N}
    \Delta(\varphi_{1,2},\ldots,\varphi_{N-1,N})=\mathcal{I},
\end{eqnarray}
where the integral extends over any $2\pi$ intervals of the form
$(\varphi_{k},\varphi_{k}+2\pi)$ and $\varphi_{k}$ are the
reference phases for all $k=1,2,\ldots,N$.
A general and symmetric POVM in a single $N_{j}$-dimensional
Hilbert space $\mathcal{H}_{\mathcal{Q}_{j}}$ is given by
\begin{eqnarray}
&&\Delta(\varphi_{1_j,2_j},\ldots,\varphi_{1_j,N_j},
\varphi_{2_j,3_j},\ldots,\varphi_{N_{j}-1,N_{j}})\\\nonumber&=&
\sum^{N_{j}}_{l_{j}}\sum^{N_{j}}_{k_{j}=1}
e^{i\varphi_{k_{j},l_{j}}}\ket{k_{j}}\bra{l_{j}},
\end{eqnarray}
where $\ket{k_{j}}$ and $\ket{l_{j}}$ are the basis vectors in
$\mathcal{H}_{\mathcal{Q}_j}$ and quantum phases satisfies the
following relation $
\varphi_{k_{j},l_{j}}-\varphi_{l_{j},k_{j}}(1-\delta_{k_{j}
l_{j}})$. The POVM is a function of the $N_{j}(N_{j}-1)/2$ phases
$(\varphi_{1_j,2_j},\ldots,\varphi_{1_j,N_j},\varphi_{2_j,3_j},\ldots,\varphi_{N_{j}-1,N_{j}})$.
It is now possible to form a POVM of a multipartite system by
simply forming the tensor product
\begin{eqnarray}\label{POVM}
&&\Delta_\mathcal{Q}(\varphi_{\mathcal{Q}_{1};k_{1},l_{1}},\ldots,
\varphi_{\mathcal{Q}_{m};k_{m},l_{m}})\\\nonumber&=&
\Delta_{\mathcal{Q}_{1}}(\varphi_{\mathcal{Q}_{1};k_{1},l_{1}})
\otimes\cdots
\otimes\Delta_{\mathcal{Q}_{m}}(\varphi_{\mathcal{Q}_{m};k_{m},l_{m}}),
\end{eqnarray}
where, e.g., $\varphi_{\mathcal{Q}_{j};k_{j},l_{j}}$ is the set of
POVMs relative phase associated with subsystems $\mathcal{Q}_{j}$,
for all $k_{j},l_{j}=1,2,\ldots,N_{j}$, where we need only to
consider when $l_{j}>k_{j}$. This POVM will play a central role in
constructing concurrence classes for multi-qubit states.

\section{Entanglement of formation and concurrence}

In this section we will review entanglement of formation and
concurrence for a pair of qubits and a general bipartite state.
For a mixed quantum system $\mathcal{Q}_{2}(N_{1},N_{2})$ the
entanglement of formation is defined by
\begin{eqnarray}
\mathcal{E}_{F}(\mathcal{Q}_{2}(N_{1},N_{2}))
&=&\inf\sum_{n}p_{n}\mathcal{E}_{F}(\rho^{p}_{\mathcal{Q}(n)}),
\end{eqnarray}
where $0\leq p_{n}\leq 1$ is a probability distribution and the
infimum is taken over all pure state decomposition of
$\rho_{\mathcal{Q}}$. The entanglement of formation for a mixed
quantum system $\mathcal{Q}_{2}(2,2)$ \cite{Wootters98} can be
written in term of the Shannon entropy and concurrence as follows
\begin{eqnarray}
\mathcal{E}_{F}(\mathcal{Q}_{2}(2,2))&=&
\mathrm{H}\left(\frac{1}{2}\left(1+\left(1-\mathcal{C}^{2}
\left(\mathcal{Q}_{2}(2,2)\right)\right)^{\frac{1}{2}}\right)\right),
\end{eqnarray}
where $\mathcal{C} \left(\mathcal{Q}_{2}(2,2)\right)$ is called
concurrence and is defined by
\begin{eqnarray}\label{eq:concx}
\mathcal{C}
\left(\mathcal{Q}_{2}(2,2)\right)=\max(0,\lambda_{1}-\sum_{n>1}\lambda_{n}),
\end{eqnarray}
where, $\lambda_{n},~n=1,...,4$ are square roots of the
eigenvalues of $\rho_{\mathcal{Q}}\tilde{\rho}_{\mathcal{Q}}$ in
descending order, where $\tilde{\rho}_{\mathcal{Q}}$ is given by $
\tilde{\rho}_{\mathcal{Q}}=(\sigma_{2}\otimes\sigma_{2})\rho^{*}_{\mathcal{Q}}(\sigma_{2}\otimes\sigma_{2})
$,
 $\mathrm{H}(\mathrm{X})$ is the Shannon entropy and $\sigma_{2}=\left(%
\begin{array}{cc}
  0 & -i \\
  i & 0 \\
\end{array}%
\right)$ is the Pauli matrix.
Moreover, the concurrence of a pure two-qubit, bipartite state is
defined as
  $\mathcal{C}(\Psi)=|\langle\Psi\ket{\widetilde{\Psi}}|$,
  where  the tilde represents the "spin-flip" operation
  $\ket{\widetilde{\Psi}}=\sigma_{2}\otimes
  \sigma_{2}\ket{\Psi^{*}}$.
In the following section we will use the concept of orthogonal
complement of our POVM to detect and to define concurrence for an
arbitrary two-qubit state and concurrence classes for arbitrary
three-, four-, and multi-qubit states.

\section{Concurrence for an arbitrary two-qubit state}
In this section we will construct concurrence for an arbitrary
two-qubit state based on orthogonal complement of  our POVM. For
two-qubit state $\mathcal{Q}_{2}(2,2)$ the POVM is explicitly
given by {\small\begin{eqnarray}
&&\Delta_\mathcal{Q}(\varphi_{\mathcal{Q}_{1};1,2},\varphi_{\mathcal{Q}_{2};1,2})=
\Delta_{\mathcal{Q}_{1}}(\varphi_{\mathcal{Q}_{1};1,2})
\otimes\Delta_{\mathcal{Q}_{2}}(\varphi_{\mathcal{Q}_{2};1,2})\\\nonumber&&=
\left(%
\begin{array}{cc}
  1 & e^{i\varphi_{\mathcal{Q}_{1}};1,2} \\
  e^{-i\varphi_{\mathcal{Q}_{1}};1,2} & 1 \\
\end{array}%
\right)\otimes\left(%
\begin{array}{cc}
  1 & e^{i\varphi_{\mathcal{Q}_{2}};1,2} \\
  e^{-i\varphi_{\mathcal{Q}_{2}};1,2} & 1 \\
\end{array}%
\right),
\end{eqnarray}}
 In this POVM, the only terms that has
information about joint properties of both subsystems are phase
sum $e^{\pm
i(\varphi_{\mathcal{Q}_{1};1,2}+\varphi_{\mathcal{Q}_{2};1,2})}$
and phase difference $e^{\pm
i(\varphi_{\mathcal{Q}_{1};1,2}-\varphi_{\mathcal{Q}_{2};1,2})}$.
Now, from this observation we can assume that the phase sum gives
a negative contribution that is $-1$ and phase difference gives a
positive contribution that is $+1$ to a measurement. Then, we can
mathematically achieve this  construction by defining an operator
$\widetilde{\Delta}_{\mathcal{Q}_{j}}(\varphi_{\mathcal{Q}_{j};1,2})=\mathcal{I}_{2}-
\Delta_{\mathcal{Q}_{j}}(\varphi_{\mathcal{Q}_{j};1,2})$, where
$\mathcal{I}_{2}$ is a 2-by-2 identity matrix, for each subsystem
$j$. Indeed by construction this operator is orthogonal complement
of our POVM. Then, we define an
 operator that detects entanglement as follows
\begin{eqnarray}
 \widetilde{\Delta}^{
EPR}_\mathcal{Q}&=&\widetilde{\Delta}_{\mathcal{Q}_{1}}(\varphi^{\frac{\pi}{2}}_{\mathcal{Q}_{1};1,2})
\otimes\widetilde{\Delta}_{\mathcal{Q}_{2}}(\varphi^{\frac{\pi}{2}}_{\mathcal{Q}_{2};1,2})\\\nonumber&=&
\sigma_{y}\otimes\sigma_{y},
\end{eqnarray}
where by choosing
$\varphi^{\frac{\pi}{2}}_{\mathcal{Q}_{j};k_{j},l_{j}}=\frac{\pi}{2}$
for all $k_{j}<l_{j}, ~j=1,2$, we get an operator which coincides
with
 Pauli spin-flip operator $\sigma_{y}$ for a single-qubit. Now, in analogy with Wootter's
formula for concurrence of a quantum system $\mathcal{Q}_{2}(2,2)$
with the density operator $\rho_{\mathcal{Q}}$, we can define
$\widetilde{\rho}^{EPR}_{\mathcal{Q}}$ as
\begin{equation}
\widetilde{\rho}^{EPR}_{\mathcal{Q}}= \widetilde{\Delta}^{
EPR}_\mathcal{Q}\rho^{*}_{\mathcal{Q}}\widetilde{\Delta}^{
EPR}_\mathcal{Q}
\end{equation}
and the  concurrence is given by $ \mathcal{C}_{\Theta}
\left(\mathcal{Q}_{2}(2,2)\right)=
\max(0,\lambda^{EPR}_{1}-\sum_{n>1}\lambda^{EPR}_{n}), $ where
$\lambda^{EPR}_{n},~n=1,\ldots,4$ are square roots of the
eigenvalues of
$\rho_{\mathcal{Q}}\widetilde{\rho}^{EPR}_{\mathcal{Q}}$ in
descending order and $\rho^{*}_{\mathcal{Q}}$ is the complex
conjugation of $\rho_{\mathcal{Q}}$.
 Now, we would like to extend this result to a three-qubit
state.
\section{Concurrence for an arbitrary three-qubit state}
The procedure of defining concurrence for an arbitrary three-qubit
state is more complicated than for a pair of qubits since in the
three-qubit state case we have to deal with two different classes
of three partite state, namely $W^{3}$ and $GHZ^{3}$ classes. For
$W^{3}$ class, we have three types of entanglement: entanglement
between subsystems one and two $\mathcal{Q}_{1}\mathcal{Q}_{2}$,
one and three $\mathcal{Q}_{1}\mathcal{Q}_{3}$, and two and three
$\mathcal{Q}_{2}\mathcal{Q}_{3}$. So there should be three
operators $\widetilde{\Delta}^{W^{3}}_{\mathcal{Q}_{1,2}}$,
$\widetilde{\Delta}^{ W^{3}}_{\mathcal{Q}_{1,3}}$ and
$\widetilde{\Delta}^{ W^{3}}_{\mathcal{Q}_{2,3}}$ corresponding to
entanglement between these subsystems, e.g., we have
 \begin{equation}
 \widetilde{\Delta}^{ W^{3}}_{\mathcal{Q}_{1,2}}=\widetilde{\Delta}_{\mathcal{Q}_{1}}
(\varphi^{\frac{\pi}{2}}_{\mathcal{Q}_{1}})
\otimes\widetilde{\Delta}_{\mathcal{Q}_{2}}
(\varphi^{\frac{\pi}{2}}_{\mathcal{Q}_{2}})\otimes
\mathcal{I}_{2},
\end{equation}
 \begin{equation}
 \widetilde{\Delta}^{ W^{3}}_{\mathcal{Q}_{1,3}}=\widetilde{\Delta}_{\mathcal{Q}_{1}}
(\varphi^{\frac{\pi}{2}}_{\mathcal{Q}_{1}})
\otimes\mathcal{I}_{2}\otimes \widetilde{\Delta}_{\mathcal{Q}_{3}}
(\varphi^{\frac{\pi}{2}}_{\mathcal{Q}_{3}}),
\end{equation}
 \begin{equation}
 \widetilde{\Delta}^{ W^{3}}_{\mathcal{Q}_{2,3}}=\mathcal{I}_{2}
\otimes\widetilde{\Delta}_{\mathcal{Q}_{2}}
(\varphi^{\frac{\pi}{2}}_{\mathcal{Q}_{2}})
\otimes\widetilde{\Delta}_{\mathcal{Q}_{3}}
(\varphi^{\frac{\pi}{2}}_{\mathcal{Q}_{3}}).
\end{equation}
Now, for a pure quantum system $\mathcal{Q}^{p}_{3}(2,2,2)$ we
define concurrence of $W^{3}$ class by
\begin{eqnarray}\label{coneW3}
    \mathcal{C}(\mathcal{Q}^{W^{3}}_{3}(2,2,2))&=&
    \left(\mathcal{N}^{W}_{3}\sum^{3}_{1=r_{1}<r_{2}}\left|\langle
    \Psi\ket{\widetilde{\Delta}^{ W^{3}}_{\mathcal{Q}_{r_{1},r_{2}}}
    \Psi^{*}}\right|^{2}\right)^{1/2}\\\nonumber&=&
    (4\mathcal{N}^{W}_{3}
[|\alpha_{1,2,1}\alpha_{2,1,1}- \alpha_{1,1,1}\alpha_{2,2,1}
\\\nonumber&&+
\alpha_{1,2,2}\alpha_{2,1,2}-\alpha_{1,1,2}\alpha_{2,2,2} |^{2}
\\\nonumber&&+ |
\alpha_{1,1,2}\alpha_{2,1,1}-\alpha_{1,1,1}\alpha_{2,1,2}\\\nonumber&&
+\alpha_{1,2,2}\alpha_{2,2,1}-\alpha_{1,2,1}\alpha_{2,2,2}
|^{2}\\\nonumber&&+|\alpha_{1,1,2}\alpha_{1,2,1}
-\alpha_{1,1,1}\alpha_{1,2,2}\\\nonumber&&+
\alpha_{2,1,2}\alpha_{2,2,1} -\alpha_{2,1,1}\alpha_{2,2,2}
|^{2}])^{1/2},
\end{eqnarray}
where $\mathcal{N}^{W}_{3}$ is a normalization constant and for a
quantum system $\mathcal{Q}_{3}(2,2,2)$ with the density operator
$\rho_{\mathcal{Q}}$, let \begin{equation}
 \widetilde{\rho}^{ W^{3}}_{\mathcal{Q}}=  \widetilde{\Delta}^{ W^{3}}_{\mathcal{Q}_{r_{1},r_{2}}} \rho^{*}_{\mathcal{Q}}
\widetilde{\Delta}^{ W^{3}}_{\mathcal{Q}_{r_{1},r_{2}}}.
\end{equation}
Then concurrence of a three-qubit mixed state of $W^{3}$ class
could be defined by
\begin{eqnarray}\label{CW3}\nonumber
\mathcal{C}(\mathcal{Q}^{W^{3}}_{3}(2,2,2))&=&
\max(0,\lambda^{W^{3}}_{1}(r_{1},r_{2})-\sum_{n>1}
\lambda^{W^{3}}_{n}(r_{1},r_{2})),
\end{eqnarray}
 where
$\lambda^{W^{3}}_{n}(r_{1},r_{2})$ for all $1\leq r_{1}<r_{2}\leq 3$
are square roots of the eigenvalues of
$\rho_{\mathcal{Q}}\widetilde{\rho}^{ W^{3}}_{\mathcal{Q}}$ in
descending order.  The second class of three-qubit state that we
would like to consider is $GHZ^{3}$ class. For $GHZ^{3}$ class we
have again three types of entanglement that give contribution to
degree of entanglement, but there is a difference in construction of
operators compare to $W^{3}$ class. The  operators
$\widetilde{\Delta}^{GHZ^{3}}_{\mathcal{Q}_{1,2}}$,
$\widetilde{\Delta}^{GHZ^{3}}_{\mathcal{Q}_{1,3}}$ and
$\widetilde{\Delta}^{GHZ^{3}}_{\mathcal{Q}_{2,3}}$ that can detect
entanglement between these subsystems, are given by
 \begin{equation}
 \widetilde{\Delta}^{GHZ^{3}}_{\mathcal{Q}_{1,2}}=\widetilde{\Delta}_{\mathcal{Q}_{1}}
(\varphi^{\frac{\pi}{2}}_{\mathcal{Q}_{1}})
\otimes\widetilde{\Delta}_{\mathcal{Q}_{2}}
(\varphi^{\frac{\pi}{2}}_{\mathcal{Q}_{2}})\otimes
\widetilde{\Delta}_{\mathcal{Q}_{3}}
(\varphi^{\pi}_{\mathcal{Q}_{3}}),
\end{equation}
 \begin{equation}
 \widetilde{\Delta}^{GHZ^{3}}_{\mathcal{Q}_{1,3}}=\widetilde{\Delta}_{\mathcal{Q}_{1}}
(\varphi^{\frac{\pi}{2}}_{\mathcal{Q}_{1}})
\otimes\widetilde{\Delta}_{\mathcal{Q}_{2}}
(\varphi^{\pi}_{\mathcal{Q}_{2}})\otimes
\widetilde{\Delta}_{\mathcal{Q}_{3}}
(\varphi^{\frac{\pi}{2}}_{\mathcal{Q}_{3}}),
\end{equation}
 \begin{equation}
\widetilde{
\Delta}^{GHZ^{3}}_{\mathcal{Q}_{2,3}}=\widetilde{\Delta}_{\mathcal{Q}_{1}}
(\varphi^{\pi}_{\mathcal{Q}_{1}})
\otimes\widetilde{\Delta}_{\mathcal{Q}_{2}}
(\varphi^{\frac{\pi}{2}}_{\mathcal{Q}_{2}})
\otimes\widetilde{\Delta}_{\mathcal{Q}_{3}}
(\varphi^{\frac{\pi}{2}}_{\mathcal{Q}_{3}}),
\end{equation}
where $\varphi^{\pi}_{\mathcal{Q}_{j}}=\pi$ for all $j$. Now, for
a pure quantum system $\mathcal{Q}^{p}_{3}(2,2,2)$ we define
concurrence of $GHZ^{3}$ class by
\begin{eqnarray}\label{cone}
    &&\mathcal{C}(\mathcal{Q}^{GHZ^{3}}_{3}(2,2,2))=\\\nonumber&&
    \left(\mathcal{N}^{GHZ}_{3}\sum^{3}_{1=r_{1}<r_{2}}
    \left|\langle \Psi\ket{\widetilde{\Delta}^{GHZ^{3}}_{\mathcal{Q}_{r_{1},r_{2}}}
\Psi^{*}}\right|^{2}\right)^{1/2},
\end{eqnarray}
where $\mathcal{N}^{GHZ}_{3}$ is a normalization constant.
For a quantum system $\mathcal{Q}_{3}(2,2,2)$, let $
 \widetilde{\rho}^{GHZ^{3}}_{\mathcal{Q}}=  \widetilde{\Delta}^{GHZ^{3}}_{\mathcal{Q}_{r_{1},r_{2}}} \rho^{*}_{\mathcal{Q}}
\widetilde{\Delta}^{GHZ^{3}}_{\mathcal{Q}_{r_{1},r_{2}}}.$ Then
concurrence of a three-qubit mixed state of $GHZ^{3}$ class is
defined by
\begin{eqnarray}\label{CW3}
\mathcal{C}(\mathcal{Q}^{GHZ^{3}}_{3}(2,2,2))&=&
\max(0,\lambda^{GHZ^{3}}_{1}(r_{1},r_{2})\\\nonumber&-&\sum_{n>1}
\lambda^{GHZ^{3}}_{n}(r_{1},r_{2})),
\end{eqnarray}
 where
$\lambda^{GHZ^{3}}_{n}(r_{1},r_{2})$ for all $1\leq r_{1}<r_{2}\leq
3$ are square roots of the eigenvalues of
$\rho_{\mathcal{Q}}\widetilde{\rho}^{GHZ^{3}}_{\mathcal{Q}}$ in
descending order. For three-qubit state the operators $\Delta^{
W^{3}}_{\mathcal{Q}_{r_{1},r_{2}}}$ and
$\Delta^{GHZ^{3}}_{\mathcal{Q}_{r_{1},r_{2}}}$ satisfies $(\Delta^{
W^{3}}_{\mathcal{Q}_{r_{1},r_{2}}})^{2}=1$ and
$(\Delta^{GHZ^{3}}_{\mathcal{Q}_{r_{1},r_{2}}})^{2}=1$.
Now, for a state
$\ket{\Psi_{\overline{W}^{3}}}=\alpha_{1,2,2}\ket{1,2,2}+\alpha_{2,1,2}\ket{2,1,2}
+\alpha_{2,2,1}\ket{2,2,1}$, the $W^{3}$ class  concurrence gives
\begin{eqnarray}\nonumber
\mathcal{C}(\mathcal{Q}^{W^{3}}_{3}(2,2,2))&=&(4\mathcal{N}^{W}_{3}
[ \left|\alpha_{1,2,2}\alpha_{2,1,2} \right|^{2}+\left |
\alpha_{1,2,2}\alpha_{2,2,1} \right|^{2}\\\nonumber&&+\left
|\alpha_{2,1,2}\alpha_{2,2,1}
 \right|^{2}])^{1/2}.
\end{eqnarray}
When
    $\alpha_{1,2,2}=\alpha_{2,1,2}=\alpha_{2,2,1}=\frac{1}{\sqrt{3}}$, we get
    $\mathcal{C}(\mathcal{Q}^{W^{3}}_{3}(2,2,2))=
    (\frac{4}{3}\mathcal{N}^{W}_{3})^{1/2}$ and $\mathcal{C}(\mathcal{Q}^{GHZ^{3}}_{3}(2,2,2))=0$.
    Thus, for $\mathcal{N}^{W}_{3}=\frac{3}{4}$, we have
    $\mathcal{C}(\mathcal{Q}^{W^{3}}_{3}(2,2,2))=1$.

Moreover, let
$\ket{\Psi^{\pm}_{GHZ^{3}}}=\alpha_{1,1,1}\ket{1,1,1}\pm\alpha_{2,2,2}
\ket{2,2,2} $  and
$\rho^{GHZ}=q\ket{\Psi^{+}_{GHZ^{3}}}\bra{\Psi^{+}_{GHZ^{3}}}+
(1-q)\ket{\Psi^{-}_{GHZ^{3}}}\bra{\Psi^{-}_{GHZ^{3}}}$. Then the
$GHZ^{3}$ concurrence class gives
$\mathcal{C}(\mathcal{Q}^{GHZ^{3}}_{3}(2,2,2))=
\max(0,\lambda^{GHZ^{3}}_{1}(r_{1},r_{2})-\sum_{2>1}
\lambda^{GHZ^{3}}_{n}(r_{1},r_{2})) =\max(0,2q-1)$, where
$\lambda^{GHZ^{3}}_{1}(1,2)=q$, $\lambda^{GHZ^{3}}_{2}(1,2)=1-q$,
and $0< q\leq 1$.

As we have seen there are $W^{3}$ and $GHZ^{3}$ classes
concurrences for three-qubit state. However, we are not sure how
we should deal with these two different classes, but there are at
least two possibilities: the first possibility is to deal with
them separately, and the second one is to define an overall
expression for concurrence of three-qubit state by adding these
two concurrences.
\section{Concurrence classes for an arbitrary four-qubit  state }\label{Sec4}
In this section we will construct three different concurrences for
four-qubit states based on quantum phases of our POVM, namely the
$W^{4}$, $GHZ^{4}$, and $GHZ^{3}$ class concurrences.
Let us begin by constructing operators for $W^{4}$ class of
four-qubit states. For $W^{4}$ class we have six different types
of entanglement: entanglement between subsystem one and two
$\mathcal{Q}_{1}\mathcal{Q}_{2}$, one and three
$\mathcal{Q}_{1}\mathcal{Q}_{3}$, and two and three
$\mathcal{Q}_{2}\mathcal{Q}_{3}$, etc.. So, there are six
 operators $\widetilde{\Delta}^{W^{4}}_{\mathcal{Q}_{1,2}}$,
 $\widetilde{\Delta}^{W^{4}}_{\mathcal{Q}_{1,3}}$,
$\widetilde{\Delta}^{W^{4}}_{\mathcal{Q}_{1,4}}$,
$\widetilde{\Delta}^{W^{4}}_{\mathcal{Q}_{2,3}}$,
$\widetilde{\Delta}^{W^{4}}_{\mathcal{Q}_{2,4}}$,
$\widetilde{\Delta}^{W^{4}}_{\mathcal{Q}_{3,4}}$ corresponding to
entanglement between these subsystems, i.e., we have
 \begin{equation}
 \widetilde{\Delta}^{
W^{4}}_{\mathcal{Q}_{1,2}}=\widetilde{\Delta}_{\mathcal{Q}_{1}}
(\varphi^{\frac{\pi}{2}}_{\mathcal{Q}_{1}})
\otimes\widetilde{\Delta}_{\mathcal{Q}_{2}}
(\varphi^{\frac{\pi}{2}}_{\mathcal{Q}_{2}})
\otimes\mathcal{I}_{2}\otimes \mathcal{I}_{2},
\end{equation}
 \begin{equation}
\widetilde{\Delta}^{
W^{4}}_{\mathcal{Q}_{1,3}}=\widetilde{\Delta}_{\mathcal{Q}_{1}}
(\varphi^{\frac{\pi}{2}}_{\mathcal{Q}_{1}}) \otimes
\mathcal{I}_{2}\otimes \widetilde{\Delta}_{\mathcal{Q}_{3}}
(\varphi^{\frac{\pi}{2}}_{\mathcal{Q}_{3}})\otimes
\mathcal{I}_{2},
\end{equation}
 \begin{equation}
 \widetilde{\Delta}^{
W^{4}}_{\mathcal{Q}_{1,4}}=\widetilde{\Delta}_{\mathcal{Q}_{1}}
(\varphi^{\frac{\pi}{2}}_{\mathcal{Q}_{1}})\otimes \mathcal{I}_{2}
\otimes \mathcal{I}_{2}\otimes\widetilde{\Delta}_{\mathcal{Q}_{4}}
(\varphi^{\frac{\pi}{2}}_{\mathcal{Q}_{4}}),
\end{equation}
 \begin{equation}
 \widetilde{\Delta}^{W^{4}}_{\mathcal{Q}_{2,3}}=\mathcal{I}_{2}
\otimes\widetilde{\Delta}_{\mathcal{Q}_{2}}
(\varphi^{\frac{\pi}{2}}_{\mathcal{Q}_{2}})
\otimes\widetilde{\Delta}_{\mathcal{Q}_{3}}
(\varphi^{\frac{\pi}{2}}_{\mathcal{Q}_{3}})\otimes
\mathcal{I}_{2},
\end{equation}
 \begin{equation}
\widetilde{ \Delta}^{ W^{4}}_{\mathcal{Q}_{2,4}}=\mathcal{I}_{2}
\otimes\widetilde{\Delta}_{\mathcal{Q}_{2}}
(\varphi^{\frac{\pi}{2}}_{\mathcal{Q}_{2}}) \otimes
\mathcal{I}_{2}\otimes\widetilde{\Delta}_{\mathcal{Q}_{4}}
(\varphi^{\frac{\pi}{2}}_{\mathcal{Q}_{4}}),
\end{equation}
 \begin{equation}
 \widetilde{\Delta}^{W^{4}}_{\mathcal{Q}_{3,4}}=\mathcal{I}_{2}\otimes \mathcal{I}_{2}\otimes
\widetilde{\Delta}_{\mathcal{Q}_{3}}
(\varphi^{\frac{\pi}{2}}_{\mathcal{Q}_{3}})
\otimes\widetilde{\Delta}_{\mathcal{Q}_{4}}
(\varphi^{\frac{\pi}{2}}_{\mathcal{Q}_{4}}).
\end{equation}
Now, for a pure quantum system $\mathcal{Q}^{p}_{4}(2,\ldots,2)$
we define concurrence of $W^{4}$ class by
\begin{eqnarray}\label{cone}
    \mathcal{C}(\mathcal{Q}^{W^{4}}_{4}(2,\ldots,2))&=&
    \left(\mathcal{N}^{W}_{4}\sum^{4}_{1=r_{1}<r_{2}}\left|\langle \Psi\ket{\widetilde{\Delta}^{
W^{4}}_{\mathcal{Q}_{r_{1},r_{2}}}\Psi^{*}}\right|^{2}\right)^{1/2},
\end{eqnarray}
where $\mathcal{N}^{W}_{4}$ is a normalization constant. Now, for a
quantum system $\mathcal{Q}^{W^{4}}_{2}(2,\ldots,2)$ let
$\widetilde{\rho}^{W^{4}}_{\mathcal{Q}}=\widetilde{\Delta}^{
W^{4}}_{\mathcal{Q}_{r_{1},r_{2}}} \rho^{*}_{\mathcal{Q}}
\widetilde{\Delta}^{W^{4}}_{\mathcal{Q}_{r_{1},r_{2}}} $. Then
concurrence of four-qubit mixed state of $W^{4}$ class can be
defined by
\begin{eqnarray}\label{CW4}
\mathcal{C}(\mathcal{Q}^{W^{4}}_{4}(2,\ldots,2))&=&
\max(0,\lambda^{W^{4}}_{1}(r_{1},r_{2})\\\nonumber&-&\sum_{n>1}
\lambda^{W^{4}}_{n}(r_{1},r_{2})),
\end{eqnarray}
 where
$\lambda^{W^{4}}_{n}(r_{1},r_{2})$ for all $1\leq r_{1}<r_{2}\leq 4$
are square roots of the eigenvalues of
$\rho_{\mathcal{Q}}\widetilde{\rho}^{W^{4}}_{\mathcal{Q}}$ in
descending order. The operators $\widetilde{\Delta}^{
W^{4}}_{\mathcal{Q}_{r_{1},r_{2}}}$ for $W^{4}$ class satisfies
$(\widetilde{\Delta}^{W^{4}}_{\mathcal{Q}_{r_{1},r_{2}}})^{2}=1$.
Now, for a state
$\ket{\Psi_{W^{4}}}=\alpha_{1,1,1,2}\ket{1,1,1,2}+\alpha_{1,1,2,1}\ket{1,1,2,1}
+\alpha_{1,2,1,1}\ket{1,2,1,1}+\alpha_{2,1,1,1}\ket{2,1,1,1}$, the
$W^{4}$ class  concurrence
 gives
\begin{eqnarray}\nonumber
&&\mathcal{C}(\mathcal{Q}^{W^{4}}_{4}(2,\ldots,2))=(4\mathcal{N}^{W}_{3}
[ \left|\alpha_{1,2,1,1}\alpha_{2,1,1,1}
\right|^{2}\\\nonumber&&+\left | \alpha_{1,1,2,1}\alpha_{2,1,1,1}
\right|^{2}+\left |\alpha_{1,1,1,2}\alpha_{2,1,1,1}
 \right|^{2}+\left
|\alpha_{1,1,2,1}\alpha_{1,2,1,1}
 \right|^{2}\\\nonumber&&+\left
|\alpha_{1,1,1,2}\alpha_{1,2,1,1}
 \right|^{2}+\left
|\alpha_{1,1,1,2}\alpha_{1,1,2,1}
 \right|^{2}]
 )^{1/2}
\end{eqnarray} and for
    $\alpha_{1,1,1,2}=\alpha_{1,1,2,1}=\alpha_{1,2,1,1}=\alpha_{1,2,1,1}
    =\frac{1}{\sqrt{4}}$, we get
    $\mathcal{C}(\mathcal{Q}^{W^{4}}_{4}(2,\ldots,2))=
    (\frac{3}{2}\mathcal{N}^{W}_{4})^{1/2}$,
    $\mathcal{C}(\mathcal{Q}^{GHZ^{3}}_{4}(2,\ldots,2))=0$.
 The second class of four-qubit state that we would like to
consider is $GHZ^{4}$ class.
For $GHZ^{4}$, we have again six different types of entanglement
and there are six  operators defined as follows
 \begin{eqnarray}
 \widetilde{\Delta}^{
GHZ^{4}}_{\mathcal{Q}_{1,2}}&=&\widetilde{\Delta}_{\mathcal{Q}_{1}}
(\varphi^{\frac{\pi}{2}}_{\mathcal{Q}_{1}})
\otimes\widetilde{\Delta}_{\mathcal{Q}_{2}}
(\varphi^{\frac{\pi}{2}}_{\mathcal{Q}_{2}})
\\\nonumber&&\otimes\widetilde{\Delta}_{\mathcal{Q}_{3}}
(\varphi^{\pi}_{\mathcal{Q}_{3}})\otimes
\widetilde{\Delta}_{\mathcal{Q}_{4}}
(\varphi^{\pi}_{\mathcal{Q}_{4}}),
\end{eqnarray}
  \begin{eqnarray}
 \widetilde{\Delta}^{
GHZ^{4}}_{\mathcal{Q}_{1,3}}&=&\widetilde{\Delta}_{\mathcal{Q}_{1}}
(\varphi^{\frac{\pi}{2}}_{\mathcal{Q}_{1}})
\otimes\widetilde{\Delta}_{\mathcal{Q}_{2}}
(\varphi^{\pi}_{\mathcal{Q}_{2}})
\\\nonumber&&\otimes\widetilde{\Delta}_{\mathcal{Q}_{3}}
(\varphi^{\frac{\pi}{2}}_{\mathcal{Q}_{3}})\otimes
\widetilde{\Delta}_{\mathcal{Q}_{4}}
(\varphi^{\pi}_{\mathcal{Q}_{4}}),
\end{eqnarray}
\begin{eqnarray}
 \widetilde{\Delta}^{
GHZ^{4}}_{\mathcal{Q}_{1,4}}&=&\widetilde{\Delta}_{\mathcal{Q}_{1}}
(\varphi^{\frac{\pi}{2}}_{\mathcal{Q}_{1}})
\otimes\widetilde{\Delta}_{\mathcal{Q}_{2}}
(\varphi^{\pi}_{\mathcal{Q}_{2}})
\\\nonumber&&\otimes\widetilde{\Delta}_{\mathcal{Q}_{3}}
(\varphi^{\pi}_{\mathcal{Q}_{3}})\otimes
\widetilde{\Delta}_{\mathcal{Q}_{4}}
(\varphi^{\frac{\pi}{2}}_{\mathcal{Q}_{4}}),
\end{eqnarray}
 \begin{eqnarray}
 \widetilde{\Delta}^{
GHZ^{4}}_{\mathcal{Q}_{2,3}}&=&\otimes\widetilde{\Delta}_{\mathcal{Q}_{1}}
(\varphi^{\pi}_{\mathcal{Q}_{1}})
\otimes\widetilde{\Delta}_{\mathcal{Q}_{2}}
(\varphi^{\frac{\pi}{2}}_{\mathcal{Q}_{2}})
\\\nonumber&&\otimes\widetilde{\Delta}_{\mathcal{Q}_{3}}
(\varphi^{\frac{\pi}{2}}_{\mathcal{Q}_{3}})\otimes
\widetilde{\Delta}_{\mathcal{Q}_{4}}
(\varphi^{\pi}_{\mathcal{Q}_{4}}),
\end{eqnarray}
  \begin{eqnarray}
 \widetilde{\Delta}^{
GHZ^{4}}_{\mathcal{Q}_{2,4}}&=&\otimes\widetilde{\Delta}_{\mathcal{Q}_{1}}
(\varphi^{\pi}_{\mathcal{Q}_{1}})
\otimes\widetilde{\Delta}_{\mathcal{Q}_{2}}
(\varphi^{\frac{\pi}{2}}_{\mathcal{Q}_{2}})
\\\nonumber&&\otimes\widetilde{\Delta}_{\mathcal{Q}_{3}}
(\varphi^{\pi}_{\mathcal{Q}_{3}})\otimes
\widetilde{\Delta}_{\mathcal{Q}_{4}}
(\varphi^{\frac{\pi}{2}}_{\mathcal{Q}_{4}}),
\end{eqnarray}
 \begin{eqnarray}
 \widetilde{\Delta}^{
GHZ^{4}}_{\mathcal{Q}_{3,4}}&=&\otimes\widetilde{\Delta}_{\mathcal{Q}_{1}}
(\varphi^{\pi}_{\mathcal{Q}_{1}})
\otimes\widetilde{\Delta}_{\mathcal{Q}_{2}}
(\varphi^{\pi}_{\mathcal{Q}_{2}})
\\\nonumber&&\otimes\widetilde{\Delta}_{\mathcal{Q}_{3}}
(\varphi^{\frac{\pi}{2}}_{\mathcal{Q}_{3}})\otimes
\widetilde{\Delta}_{\mathcal{Q}_{4}}
(\varphi^{\frac{\pi}{2}}_{\mathcal{Q}_{4}}),
\end{eqnarray}
Now, for a pure four-qubit state $\mathcal{Q}^{p}_{4}(2,\ldots,2)$
we define concurrence of $GHZ^{4}$ class by
\begin{eqnarray}\label{coneGHZ4}
    &&\mathcal{C}(\mathcal{Q}^{GHZ^{4}}_{4}(2,\ldots,2))=\\\nonumber&&
    \left(\mathcal{N}^{GHZ}_{4}\sum^{4}_{1=r_{1}<r_{2}}\left|\langle \Psi\ket{\widetilde{\Delta}^{
GHZ^{4}}_{\mathcal{Q}_{r_{1},r_{2}}}\Psi^{*}}\right|^{2}\right)^{1/2},
\end{eqnarray}
where $\mathcal{N}^{GHZ}_{4}$ is a normalization constant and for
a
 quantum system $\mathcal{Q}^{GHZ^{4}}_{2}(2,\ldots,2)$
with $\widetilde{\rho}^{GHZ^{4}}_{\mathcal{Q}}=\widetilde{\Delta}^{
GHZ^{4}}_{\mathcal{Q}_{r_{1},r_{2}}} \rho^{*}_{\mathcal{Q}}
\widetilde{\Delta}^{GHZ^{4}}_{\mathcal{Q}_{r_{1},r_{2}}} $, we
define concurrence of four-qubit mixed state of $GHZ^{4}$ class by
\begin{eqnarray}\label{CGHZ4}
\mathcal{C}(\mathcal{Q}^{GHZ^{4}}_{4}(2,\ldots,2))&=&
\max(0,\lambda^{GHZ^{4}}_{1}(r_{1},r_{2})\\\nonumber&-&\sum_{n>1}
\lambda^{GHZ^{4}}_{n}(r_{1},r_{2})),
\end{eqnarray}
 where
$\lambda^{GHZ^{4}}_{n}(r_{1},r_{2})$ for all $1\leq r_{1}<r_{2}\leq
4$ are square roots of the eigenvalues of
$\rho_{\mathcal{Q}}\widetilde{\rho}^{GHZ^{4}}_{\mathcal{Q}}$ in
descending order. Moreover, we have $(\widetilde{\Delta}^{
GHZ^{4}}_{\mathcal{Q}_{r_{1},r_{2}}})^{2}= 1$.
The third class of four-qubit state that we want to consider is
$GHZ^{3}$ class. For $GHZ^{3}$, we have four different types of
entanglement. So there are four operators defined as below
 \begin{equation}
 \widetilde{\Delta}^{GHZ^{3}}_{\mathcal{Q}_{12,3}}=\widetilde{\Delta}_{\mathcal{Q}_{1}}
(\varphi^{\frac{\pi}{2}}_{\mathcal{Q}_{1}})
\otimes\widetilde{\Delta}_{\mathcal{Q}_{2}}
(\varphi^{\frac{\pi}{2}}_{\mathcal{Q}_{2}})\otimes
\widetilde{\Delta}_{\mathcal{Q}_{3}}
(\varphi^{\pi}_{\mathcal{Q}_{3}}) \otimes\mathcal{I}_{2},
\end{equation}
 \begin{equation}
 \widetilde{\Delta}^{GHZ^{3}}_{\mathcal{Q}_{12,4}}=\widetilde{\Delta}_{\mathcal{Q}_{1}}
(\varphi^{\frac{\pi}{2}}_{\mathcal{Q}_{1}}) \otimes
\widetilde{\Delta}_{\mathcal{Q}_{2}}
(\varphi^{\frac{\pi}{2}}_{\mathcal{Q}_{2}})\otimes
\mathcal{I}_{2}\otimes \widetilde{\Delta}_{\mathcal{Q}_{4}}
(\varphi^{\pi}_{\mathcal{Q}_{4}}),
\end{equation}
 \begin{equation}
\widetilde{\Delta}^{GHZ^{3}}_{\mathcal{Q}_{13,4}}=\widetilde{\Delta}_{\mathcal{Q}_{1}}
(\varphi^{\frac{\pi}{2}}_{\mathcal{Q}_{1}})\otimes\mathcal{I}_{2}
\otimes\widetilde{\Delta}_{\mathcal{Q}_{3}}
(\varphi^{\frac{\pi}{2}}_{\mathcal{Q}_{3}})\otimes
\widetilde{\Delta}_{\mathcal{Q}_{4}}
(\varphi^{\pi}_{\mathcal{Q}_{4}}),
\end{equation}
\begin{equation}
 \widetilde{\Delta}^{GHZ^{3}}_{\mathcal{Q}_{23,4}}=\mathcal{I}_{2}
\otimes\widetilde{\Delta}_{\mathcal{Q}_{2}}
(\varphi^{\frac{\pi}{2}}_{\mathcal{Q}_{2}})
\otimes\widetilde{\Delta}_{\mathcal{Q}_{3}}
(\varphi^{\pi}_{\mathcal{Q}_{3}})
\otimes\widetilde{\Delta}_{\mathcal{Q}_{4}}
(\varphi^{\frac{\pi}{2}}_{\mathcal{Q}_{4}}).
\end{equation}
Then, for a pure four-qubit state
$\mathcal{Q}^{p}_{4}(2,\ldots,2)$ we define concurrence for a
$GHZ^{3}$ class by
\begin{eqnarray}\label{coneGHZ3}
   && \mathcal{C}(\mathcal{Q}^{GHZ^{3}}_{4}(2,\ldots,2))=\\\nonumber&&
    \left(\mathcal{N}^{GHZ}_{3}\sum^{4}_{1=r_{1}<r_{2}<r_{3}}
    \left|\langle \Psi\ket{\widetilde{\Delta}^{GHZ^{3}}_{\mathcal{Q}_{r_{1}r_{2},r_{3}}}\Psi^{*}}\right|^{2}\right)^{1/2},
\end{eqnarray}
where $\mathcal{N}^{GHZ}_{3}$ is a normalization constant and for
a  quantum system $\mathcal{Q}_{2}(2,\ldots,2)$ with density
operator $\rho_{\mathcal{Q}}$, let
\begin{equation}
 \widetilde{\rho}^{GHZ^{3}}_{\mathcal{Q}}=  \widetilde{\Delta}^{GHZ^{3}}_{\mathcal{Q}_{r_{1}r_{2},r_{3}}} \rho^{*}_{\mathcal{Q}}
\widetilde{\Delta}^{GHZ^{3}}_{\mathcal{Q}_{r_{1}r_{2},r_{3}}}.
\end{equation}
Then concurrence  for a four-qubit $GHZ^{3}$ class is defined by
\begin{eqnarray}\label{CGHZ34}\nonumber
\mathcal{C}(\mathcal{Q}^{GHZ^{3}}_{4}(2,\ldots,2))&=&
\max(0,\lambda^{GHZ^{3}}_{1}(r_{1}r_{2},r_{3})\\&-&\sum_{n>1}
\lambda^{GHZ^{3}}_{n}(r_{1}r_{2},r_{3})),
\end{eqnarray}
 where
$\lambda^{GHZ^{3}}_{n}(r_{1}r_{2},r_{3})$ for all $1\leq
r_{1}<r_{2}<r_{3}\leq 4$ are square roots of the eigenvalues of
$\rho_{\mathcal{Q}}\widetilde{\rho}^{GHZ^{3}}_{\mathcal{Q}}$ in
descending order. And again we have
$(\Delta^{GHZ^{3}}_{\mathcal{Q}_{r_{1}r_{2},r_{3}}})^{2}=1$. Thus,
we have detected and defined three different concurrences for
four-qubit state based on our POVM construction.
\section{Concurrence classes for an arbitrary multi-qubit state}
At this point, we can realize that, in principle, we could in a
straightforward manner extend our construction into a multi-qubit
state $\mathcal{Q}_{m}(2,\ldots,2)$.  In order to simplify our
presentation,  we will use $\Lambda_{m}=k_{1},l_{1};$
$\ldots;k_{m},l_{m}$ as an abstract multi-index notation, where
$k_{j}=1,l_{j}=2$ for all $j$. The unique structure of our POVM
enables us to distinguish different classes of multipartite
states, which are inequivalent under LOCC operations. In the
$m$-partite case, the off-diagonal elements of the matrix
corresponding to
\begin{eqnarray}
&&\widetilde{\Delta}_\mathcal{Q}(\varphi_{\mathcal{Q}_{1};k_{1},l_{1}},\ldots,
\varphi_{\mathcal{Q}_{m};k_{m},l_{m}})=\\\nonumber&&
\widetilde{\Delta}_{\mathcal{Q}_{1}}(\varphi_{\mathcal{Q}_{1};k_{1},l_{1}})
\otimes\cdots
\otimes\widetilde{\Delta}_{\mathcal{Q}_{m}}(\varphi_{\mathcal{Q}_{m};k_{m},l_{m}}),
\end{eqnarray}
 have phases that are sum
or differences of phases originating from two and $m$ subsystems.
That is, in the later case the phases of
$\widetilde{\Delta}_\mathcal{Q}(\varphi_{\mathcal{Q}_{1};k_{1},l_{1}},\ldots,
\varphi_{\mathcal{Q}_{m};k_{m},l_{m}})$ take the form
$(\varphi_{\mathcal{Q}_{1};k_{1},l_{1}}\pm\varphi_{\mathcal{Q}_{2};k_{2},l_{2}}
\pm\ldots\pm\varphi_{\mathcal{Q}_{m};k_{m},l_{m}})$ and
identification of these joint phases makes our classification
possible. Thus, we  can define linear operators for  the
$EPR_{\mathcal{Q}_{r_{1}}\mathcal{Q}_{r_{2}}}$ class based on our
POVM which are sum and difference of phases of two subsystems,
i.e., $(\varphi_{\mathcal{Q}_{r_{1}};k_{r_{1}},l_{r_{1}}}
\pm\varphi_{\mathcal{Q}_{r_{2}};k_{r_{2}},l_{r_{2}}})$. That is,
for the $EPR_{\mathcal{Q}_{r_{1}}\mathcal{Q}_{r_{2}}}$  class we
have
\begin{eqnarray}\nonumber
 \widetilde{\Delta}^{
EPR_{\Lambda_{m}}}_{\mathcal{Q}_{r_{1},r_{2}}(2_{r_{1}},2_{r_{2}})}
&=&\mathcal{I}_{2_{1}} \otimes\cdots
\otimes\widetilde{\Delta}_{\mathcal{Q}_{r_{1}}}
(\varphi^{\frac{\pi}{2}}_{\mathcal{Q}_{r_{1}};k_{r_{1}},l_{r_{1}}})\\\nonumber&&
\otimes\cdots\otimes \widetilde{\Delta}_{\mathcal{Q}_{r_{2}}}
(\varphi^{\frac{\pi}{2}}_{\mathcal{Q}_{r_{2}};k_{r_{2}},l_{r_{2}}})\\&&\otimes\cdots\otimes
\mathcal{I}_{2_{m}}.
\end{eqnarray}
Let $C(m,k)=\left(%
\begin{array}{c}
  m \\
  k \\
\end{array}%
\right)$ denotes the binomial coefficient. Then there is $C(m,2)$
linear operators for the
$EPR_{\mathcal{Q}_{r_{1}}\mathcal{Q}_{r_{2}}}$  class and the set
of these operators gives the $W^{m}$ class concurrence.

 For the $GHZ^{m}$ class, we define the linear
operators based on our POVM which are sum and difference of phases
of $m$-subsystems, i.e.,
$(\varphi_{\mathcal{Q}_{r_{1}};k_{r_{1}},l_{r_{1}}}
\pm\varphi_{\mathcal{Q}_{r_{2}};k_{r_{2}},l_{r_{2}}}\pm
\ldots\pm\varphi_{\mathcal{Q}_{m};k_{m},l_{m}})$. That is, for the
$GHZ^{m}$ class we have
\begin{eqnarray}
\widetilde{\Delta}^{
GHZ^{m}_{\Lambda_{m}}}_{\mathcal{Q}_{r_{1},r_{2}}(2_{r_{1}},2_{r_{2}})}
&=&\widetilde{\Delta}_{\mathcal{Q}_{r_{1}}}
(\varphi^{\frac{\pi}{2}}_{\mathcal{Q}_{r_{1}};k_{r_{1}},l_{r_{1}}})
\otimes\\\nonumber&&\widetilde{\Delta}_{\mathcal{Q}_{r_{2}}}
(\varphi^{\frac{\pi}{2}}_{\mathcal{Q}_{r_{2}};k_{r_{2}},l_{r_{2}}})
\otimes\\\nonumber&&\widetilde{\Delta}_{\mathcal{Q}_{r_{3}}}
(\varphi^{\pi}_{\mathcal{Q}_{r_{3}};k_{r_{3}},l_{r_{3}}})
\otimes\cdots
\otimes\\\nonumber&&\widetilde{\Delta}_{\mathcal{Q}_{m}}
(\varphi^{\pi}_{\mathcal{Q}_{m-1};k_{r_{m}},l_{r_{m}}}),
\end{eqnarray}
where by choosing
$\varphi^{\pi}_{\mathcal{Q}_{j};k_{j},l_{j}}=\pi$ for all
$k_{j}<l_{j}, ~j=1,2,\ldots,m$, we get an operator which has the
structure of Pauli operator $\sigma_{x}$ embedded in a
higher-dimensional Hilbert space and coincides with $\sigma_{x}$
for a single-qubit. There are $C(m,2)$ linear operators for the
$GHZ^{m}$ class and the set of these operators gives the $GHZ^{m}$
class concurrence.

Moreover, we define the linear operators for the $GHZ^{m-1}$ class
of $m$-partite states based on our POVM which are sum and
difference of phases of $m-1$-subsystems, i.e.,
$(\varphi_{\mathcal{Q}_{r_{1}};k_{r_{1}},l_{r_{1}}}
\pm\varphi_{\mathcal{Q}_{r_{2}};k_{r_{2}},l_{r_{2}}}
\pm\ldots\varphi_{\mathcal{Q}_{m-1};k_{m-1},l_{m-1}}\pm\varphi_{\mathcal{Q}_{m-1};k_{m-1},l_{m-1}})$.
That is, for the $GHZ^{m-1}$ class we have

\begin{eqnarray}
\widetilde{\Delta}^{
GHZ^{m-1}_{\Lambda_{m}}}_{\mathcal{Q}_{r_{1}r_{2},r_{3}}(2_{r_{1}},2_{r_{2}})}
&=& \widetilde{\Delta}_{\mathcal{Q}_{r_{1}}}
(\varphi^{\frac{\pi}{2}}_{\mathcal{Q}_{r_{1}};k_{r_{1}},l_{r_{1}}})
\otimes\\\nonumber&&\widetilde{\Delta}_{\mathcal{Q}_{r_{2}}}
(\varphi^{\frac{\pi}{2}}_{\mathcal{Q}_{r_{2}};k_{r_{2}},l_{r_{2}}})
\otimes\\\nonumber&&\widetilde{\Delta}_{\mathcal{Q}_{r_{3}}}
(\varphi^{\pi}_{\mathcal{Q}_{r_{3}};k_{r_{3}},l_{r_{3}}})
\otimes\cdots
\otimes\\\nonumber&&\widetilde{\Delta}_{\mathcal{Q}_{m-1}}
(\varphi^{\pi}_{\mathcal{Q}_{m-1};k_{r_{m-1}},l_{r_{m-1}}})\otimes\mathcal{I}_{2_{m}}
,
\end{eqnarray}
where $1\leq r_{1}<r_{2}<\cdots<r_{m-1}<m$. There is $C(m,m-1)$
such operators for the $GHZ^{m-1}$ class. Now, for pure quantum
system $\mathcal{Q}^{p}_{3}(2,\ldots,2)$ we define the
$EPR_{\mathcal{Q}_{r_{1}}\mathcal{Q}_{r_{2}}}$ class concurrence
as
\begin{eqnarray}\nonumber\label{conEPR}
   &&\mathcal{C}(\mathcal{Q}^{EPR_{r_{1},r_{2}}}_{m}(2,\ldots,2))
=( 4\mathcal{N}^{EPR_{r_{1},r_{2}}}_{m}\\\nonumber&& \sum_{
k_{1},l_{1},\ldots,;k_{m},l_{m}}
    \left|\langle \Psi\ket{\widetilde{\Delta}^{
EPR_{\Lambda_{m}}}_{\mathcal{Q}_{r_{1},r_{2}}(2_{r_{1}},2_{r_{2}})}
\Psi^{*}}\right|^{2})^{1/2},
\end{eqnarray}
and the $W^{m}$ class concurrence as
\begin{eqnarray}
&&\mathcal{C}(\mathcal{Q}^{W^{m}}_{m}(2,\ldots,2))\\\nonumber&&\label{Wm}=
    \left(\sum^{m}_{r_{2}>r_{1}=1}
    \mathcal{C}^{2}(\mathcal{Q}^{EPR_{r_{1},r_{2}}}_{m}(2,\ldots,2))
    \right)^{1/2},
\end{eqnarray}
where $\mathcal{N}^{EPR_{r_{1},r_{2}}}_{m}$ are normalization
constants. Moreover, the $GHZ^{m}$ class concurrence for general
pure quantum system $\mathcal{Q}^{p}_{m}(2,\ldots,2)$ with
\begin{eqnarray}
  && \mathcal{C}(\mathcal{Q}^{GHZ^{m}}_{r_{1},r_{2}}(2_{r_{1}},2_{r_{2}}))\\\nonumber&&
    =\sum_{\forall k_{1},l_{1},\ldots,k_{m},l_{m}}
    \left|\langle \Psi\ket{\widetilde{\Delta}^{
GHZ^{m}_{\Lambda_{m}}}_{\mathcal{Q}_{r_{1},r_{2}}
(2_{r_{1}},2_{r_{2}})}\Psi^{*}} \right|^{^{2}},
\end{eqnarray}
is given by
\begin{eqnarray}\label{conGHZ}
&&\mathcal{C}(\mathcal{Q}^{GHZ^{m}}_{m}(2,\ldots,2))\\\nonumber&&=
    \left(\mathcal{N}^{GHZ}_{m}\sum^{m}_{r_{2}>r_{1}=1}\mathcal{C}(\mathcal{Q}^{GHZ^{m}}_{r_{1},r_{2}}
    (2_{r_{1}},2_{r_{2}}))\right)^{1/2},
\end{eqnarray}
where $\mathcal{N}^{GHZ}_{m}$ is a normalization constant.
 Now, let us address the monotonicity of these concurrence
classes of multipartite states. For $m$-qubit states, the $W^{m}$
class concurrences are entanglement monotones.  Let $A_{j}\in
SL(2,\mathbf{C})$, for $j=1,2,\ldots,m$, and
    $\mathcal{A}=A_{1}\otimes A_{2}\otimes\cdots\otimes A_{m}$, then
    $ \mathcal{A}\widetilde{\Delta}^{
W^{m}_{1,2;\ldots;1,2}}_{\mathcal{Q}_{r_{1},r_{2}}(2_{r_{1}},2_{r_{2}})}
\mathcal{A}^{T}=\widetilde{\Delta}^{
W^{m}_{1,2;\ldots;1,2}}_{\mathcal{Q}_{r_{1},r_{2}}(2_{r_{1}},2_{r_{2}})}$,
for all $1<r_{1}<r_{2}<m$. Thus, the $W^{m}$ class concurrences
for multi-qubit states are invariant under SLOCC, and hence are
entanglement monotones. Again, for general multipartite states we
cannot give any proof on invariance of $W^{m}$ class concurrence
under  SLOCC and this question needs further investigation.
Moreover, for multipartite states, the $GHZ^{m}$ class
concurrences are not entanglement monotone except under additional
conditions. Since
    $ \mathcal{A}\widetilde{\Delta}^{
GHZ^{m}_{1,2;\ldots;1,2}}_{\mathcal{Q}_{r_{1},r_{2}}(2_{r_{1}},2_{r_{2}})}
\mathcal{A}^{T}\neq\widetilde{\Delta}^{
GHZ^{3}_{1,2;\ldots;1,2}}_{\mathcal{Q}_{r_{1},r_{2}}(2_{r_{1}},2_{r_{2}})}$,
for all $1<r_{1}<r_{2}<m$. The reason is that $ A_{j}
\widetilde{\Delta}_{\mathcal{Q}_{j}}(\varphi^{\pi}_{\mathcal{Q}_{j};1,2})A^{T}_{j}
\neq\widetilde{\Delta}_{\mathcal{Q}_{j}}(\varphi^{\pi}_{\mathcal{Q}_{j};1,2})$.
Thus, the  $GHZ^{m}$ class concurrence for three-qubit states are
not invariant under SLOCC, and hence are not entanglement
monotones. However, by construction  the $GHZ^{m}$ class
concurrences are invariant under all permutations. Moreover, we
have $(\widetilde{\Delta}^{
GHZ^{3}_{1,2;\ldots;1,2}}_{\mathcal{Q}_{r_{1},r_{2}}(2_{r_{1}},2_{r_{2}})})^{2}=1$
and
$(\widetilde{\Delta}_{\mathcal{Q}_{j}}(\varphi^{\pi}_{\mathcal{Q}_{j};1,2}))^{2}=1$.
Furthermore, we need to be very careful when we are using the
$GHZ^{m}$ class concurrences. This class can be zero even for an
entangled multipartite state. Since we have more than two joint
phases in our POVM for $GHZ^{m}$ class concurrence. Thus, for the
$GHZ^{m}$ class concurrences we need to perform an optimization
over local unitary operations. For example, let
$\mathcal{U}=U_{1}\otimes U_{2}\otimes\cdots\otimes U_{m}$, where
$U_{j}\in U(2,\mathbf{C})$. Then we maximize the $GHZ^{m}$ class
concurrences for a given pure $m$-partite state over all local
unitary operations $\mathcal{U}$.

Finally, e.g., for the $W^{m}$ class for a general quantum system
$\mathcal{Q}_{m}(2,\ldots,2)$ with density operator
$\rho_{\mathcal{Q}}$, we define
\begin{equation}
 \widetilde{\rho}^{W^{m}_{\Lambda_{m}}}_{\mathcal{Q}}=  \widetilde{\Delta}^{
EPR_{\Lambda_{m}}}_{\mathcal{Q}_{r_{1},r_{2}}(2_{r_{1}},2_{r_{2}})}
\rho^{*}_{\mathcal{Q}} \widetilde{\Delta}^{
EPR_{\Lambda_{m}}}_{\mathcal{Q}_{r_{1},r_{2}}(2_{r_{1}},2_{r_{2}})}
\end{equation}
and then the $W^{m}$ class concurrence is defined by
\begin{eqnarray}\label{CWm}
&&\mathcal{C}(\mathcal{Q}^{W^{m}_{\Lambda_{m}}}_{2}(2,\ldots,2))\\\nonumber&=&
\max(0,\lambda^{W^{m}_{\Lambda_{m}}}_{1}(r_{1},r_{2})-\sum_{n>1}
\lambda^{W^{m}_{\Lambda_{m}}}_{n}(r_{1},r_{2})),
\end{eqnarray}
 where
$\lambda^{W^{m}_{\Lambda_{m}}}_{n}(r_{1},r_{2})$ for all $1\leq
r_{1}<r_{2}\leq m$ are the square roots of the eigenvalues of
$\rho_{\mathcal{Q}}\widetilde{\rho}^{W^{m}_{\Lambda_{m}}}_{\mathcal{Q}}$
in descending order. The $GHZ^{m}$ class concurrences for a
quantum system $\mathcal{Q}_{m}(N2,\ldots,2)$ can be defined in
similar way. The definition of concurrence classes for
multi-partite mixed states is only a well motivated suggestion and
is a generalization of Wootters and Uhlmann definitions. Moreover,
our operators $\Delta^{ X_{m}}_{\mathcal{Q}_{r_{1},r_{2}}}$
satisfies $(\Delta^{X_{m}}_{\mathcal{Q}_{r_{1},r_{2}}})^{2}=1$. As
an example of multi-qubit state let us consider a state
$\ket{W^{m}}=\frac{1}{\sqrt{m}}(\ket{1,1,\ldots,1,2}
+\ldots+\ket{2,1,\ldots,1,1})$. For this state the $W^{m}$ class
concurrence is
\begin{eqnarray}
\mathcal{C}(\mathcal{Q}^{W^{m}}_{m}(2,\ldots,2))&=&
    (\frac{4C(m,2)}{m^{2}}\mathcal{N}^{W}_{m})^{1/2}
    \\\nonumber&=&(\frac{ 2(m-1)}{m}\mathcal{N}^{W}_{m})^{1/2}.
\end{eqnarray}
 This value coincides with the one given by D\"{u}r
\cite{Dur00}. Finally, for some partially separable states the
$\mathcal{C}(\mathcal{Q}^{W^{m}}_{m}(2,\ldots,2))$ class and
$\mathcal{C}(\mathcal{Q}^{GHZ^{m}}_{m}(2,\ldots,2))$ class
concurrences do not exactly quantify entanglement in general.
Example of such states can be e.g., constructed for three-qubit
states. Thus, we may need to define a overall concurrence by
adding these concurrence classes.
\section{Conclusion}

In this paper we have expressed concurrence for an arbitrary
two-qubit state, based on our POVM, which coincides with Wootters
original formula. Moreover, we have generalized this result into
arbitrary three- and four-qubit  states. For three-qubit states, we
have found two different concurrence classes and for four-qubit
states, we have  constructed three concurrence classes. And finally,
we have generalized our result into arbitrary multi-qubit state and
we have explicitly constructed $W^{m}$ and $GHZ^{m}$ class
concurrences. We have investigate the monotonicity of the $W^{m}$
class and the $GHZ^{m}$ class concurrences for multi-qubit states.
The $W^{m}$ class concurrence for multi-qubit states are
entanglement monotones. However, $GHZ^{m}$ class concurrences need
optimization over all local unitary operation. Our construction
suggested the existence of different classes of multipartite
entanglement which are in equivalent under LOCC. At least, we known
that there is two different classes of entanglement for multi-qubit
states which our methods could distinguish very well. But we can
also define an overall expression for concurrence with a suitable
normalization coefficient. However, we think that this work is a
timely contribution to the relatively large effort presently being
undertaken to quantify and classify multipartite entanglement.

\begin{acknowledgments}
The author acknowledge useful discussions with Gunnar Bj\"{o}rk
and Piotr Badziag. The author also would like to thank Jan
Bogdanski. This work was supported by the Wenner-Gren Foundations.
\end{acknowledgments}

\end{document}